\documentclass[useAMS,usenatbib,usegraphicx]{mn2e}

\usepackage{amsfonts}
\usepackage{verbatim}
\usepackage{graphicx}

\def\lesssim{\mathrel{\hbox{\rlap{\hbox{\lower3pt\hbox{$\sim$}}}\hbox{\raise2pt\hbox{$<$}}}}}
\def\gtrsim{\mathrel{\hbox{\rlap{\hbox{\lower3pt\hbox{$\sim$}}}\hbox{\raise2pt\hbox{$>$}}}}}

\title[X-Ray Sources in Nearby Galaxies]{Searching for X-ray sources in nearby late-type galaxies with low star formation rates\thanks{Based on observations obtained with \emph{XMM-Newton}, an ESA science mission with instruments and contributions directly funded by ESA Member States and NASA.}}
\author[Chatterjee et al.]{K. Chatterjee,$^{1}$ P. Kaaret,$^{2}\thanks{E-mail: philip-kaaret@uiowa.edu}$ M. Brorby,$^{2}$ J.~J.~E. Kajava,$^{3}$ F. Gris\'e,$^{4}$ S. Farrell$^{5}$  and \newauthor J. Poutanen$^{6}$\\
$^{1}$Department of Physics, Indian Institute of Technology, Kharagpur, WB 721302, India\\
$^{2}$Department of Physics and Astronomy, University of Iowa, Iowa City, IA 52242, USA\\
$^{3}$European Space Astronomy Centre (ESA/ESAC), Science Operations Department, 28691 Villanueva de la Ca\~{n}ada, Madrid, Spain\\
$^{4}$Observatoire astronomique de Strasbourg, Universit$\acute{e}$ de Strasbourg, CNRS, UMR 7550, 11 rue de l'Universit$\acute{e}$, F-67000 Strasbourg, France \\
$^{5}$Sydney Institute for Astronomy, School of Physics, The University of Sydney, NSW 2006, Australia \\
$^{6}$Tuorla Observatory, Department of Physics and Astronomy, University of Turku, V\"{a}is\"{a}l\"{a}ntie 20, FIN-21500 P\"{u}kki\"{o}, Finland}

\begin{document}

\date{Accepted Year Month Date. Received Year Month Date; in original form Year Month Date}

\pagerange{\pageref{firstpage}--\pageref{lastpage}} \pubyear{2015}

\maketitle

\label{firstpage}

\begin{abstract}

Late type non-starburst galaxies have been shown to contain X-ray emitting objects, some being ultraluminous X-ray sources. We report on \emph{XMM-Newton} observations of 11 nearby, late-type galaxies previously observed with the \emph{Hubble Space Telescope (HST)} in order to find such objects.  We found 18 X-ray sources in or near the optical extent of the galaxies, most being point-like.  If associated with the corresponding galaxies, the source luminosities range from $2 \times 10^{37}$~erg~s$^{-1}$ to $6 \times 10^{39}$~erg~s$^{-1}$.  We found one ultraluminous X-ray source, which is in the galaxy IC~5052, and one source coincident with the galaxy IC~4662 with a blackbody temperature of $0.166 \pm 0.015$~keV that could be a quasi-soft source or a quiescent neutron star X-ray binary in the Milky Way.  One X-ray source, XMMU J205206.0$-$691316, is extended and coincident with a galaxy cluster visible on an \emph{HST} image. The X-ray spectrum of the cluster reveals a redshift of $z = 0.25 \pm 0.02$ and a temperature of 3.6$\pm$0.4 keV. The redshift was mainly determined by a cluster of Fe~XXIV lines between the observed energy range $0.8-1.0$~keV.

\end{abstract}


\begin{table*}
\centering
 \begin{minipage}{185mm}\scriptsize
 \caption{Description of the \emph{XMM-Newton} observations. Distances are redshift-independent measurements using the tip of the red giant branch (TRGB) method \citep[e.g.,][]{makarov2006} whose values are taken from the NASA/IPAC Extragalactic Datadase (NED). The live times for each detector differ from the nominal observation time and vary with respect to each other due to background flaring effects.} \label{tab:observations}
\begin{tabular}{@{}ccccccccccccc@{}}
\hline 
\hline
Obs. I.D. & Target & Type$^\dagger$&Dist. & SFR$^{\dagger\dagger}$ & \multicolumn{2}{c}{Position}&Exposure &Obs.&\multicolumn{3}{c}{Live Times after}\\
&&&(Mpc)& ($10^{-3}~{\rm M_{\sun} yr}^{-1})$ &RA&DEC&Start&Time&\multicolumn{3}{c}{filtering on RATE(ks)}\\
&&&&&(deg)&(deg)&&(ks)&MOS1&MOS2&PN \\
\hline
\hline
0721910101&UGCA 442	&SB(s)m&4.27		& 5	&355.939793&$-$31.956769&2013-12-25&19.9 &18.4&18.4&7.8\\
0721910201&NGC 784&SBdm&5.19			&36	&30.320543&+28.837258&2013-08-10&18.0 &10.5&10.4&4.8\\
0721910301&NGC 4605	&SB(s)c pec&5.47	&135&189.997416&+61.609167&2013-11-24&8.0 &5.6&5.7&0.5\\
0721910401&ESO 154$-$G023&SB(s)m&5.76	&29	&44.209999&$-$54.571389&2013-08-16&18.7 &6.7&6.8&2.8\\
0721910501&IC 5052&SBd&6.03				&87	&313.026251&$-$69.203611&2013-10-01&19.5 &18.0&17.9&14.6\\
0721910601&IC 3104&IB(s)m&2.27			&3	&184.692084&$-$79.725833&2013-09-13&11.0 &5.8&5.2&0.7\\
0721910701&IC 4662&IBm&2.44				&53	&266.793750&$-$64.640555&2013-09-26&12.0 &10.5&10.5&8.1\\
0721910801&ESO 383$-$G087&SB(s)dm&3.45	&27	&207.322875&$-$36.063422&2013-08-20&16.0 &14.4&14.3&7.1\\
0721910901&NGC 5264	&IB(s)m&4.50		&7	&205.402833&$-$29.913111&2014-01-26&8.0 &6.5&6.5&4.2\\
0721911001&NGC 1311	&SB(s)m&5.45		&20	&50.028958&$-$52.185489&2013-08-08&14.8 &8.5&8.3&4.6\\
0721911101&IC 1959&SB(s)m&6.06			&25	&53.302416&$-$50.414194&2013-08-14&14.0 &3.0&3.1&0.0\\
\hline
\end{tabular}\\
$^\dagger$ Galaxy morphology types were taken from NED which mainly utilizes the Third Reference Catalogue of Bright Galaxies, Version 3.9 (RC3.9). \\
$^{\dagger\dagger}$SFR values were estimated using H$\alpha$ luminosities taken from \citet{b14}.
\end{minipage}
\end{table*}

\section{Introduction}

The study of external galaxies enables us to probe populations of X-ray sources absent or rare in the Milky Way. The ultraluminous X-ray sources (ULXs), which represent X-ray binaries having X-ray luminosities exceeding the Eddington luminosity for a 20 M$_\odot$ compact object $(L_{\rm X} \gtrsim 2\times 10^{39}$~erg~s$^{-1})$, belong to such a class. Previous studies have shown that ULXs are likely binary systems containing either an intermediate-mass black hole (IMBH) accreting at sub-Eddington rates \citep{b16,b17,b18,Farrell09,feng_m82,pasham14} or a stellar mass black hole or neutron star with a luminosity exceeding the Eddington limit \citep{b11,b13,b19,b1,b15,motch2014}. Studying ULXs is, perhaps, the best available avenue to understand IMBHs and super-Eddington accretion.

We selected a sample of 11 nearby late-type galaxies that had all been observed with the Hubble Space Telescope (\emph{HST}), but that had not been previously observed with the Chandra X-Ray Observatory, the X-ray Multi-Mirror Mission (\emph{XMM-Newton}), or the \emph{Swift} observatory.  The \emph{HST} observations of the galaxies in our sample have yielded accurate, redshift-independent distance measurements, using the tip of the red giant branch (TRGB) method \citep[e.g.,][]{makarov2006}, that enables accurate calculation of source luminosities.  Our survey sampled mostly late-type galaxies with low star formation rates.  We obtained \emph{XMM-Newton} observations of each galaxy in order to characterize the X-ray binary population and search for ULXs. We used source detection algorithms to find X-ray sources within our \emph{XMM-Newton} images and cross-checked these against possible optical counterparts in \emph{HST} and/or X-ray counterparts in \emph{Chandra}, thereby improving the astrometry between these instruments. For the sources that are possibly associated with our target galaxies we provide photometric flux calculations and for the brightest of these we extract spectra and fit simple models. 

In Section~2, we describe the observations and analysis. In Section~3, we discuss the results of our search and describe three newly discovered X-ray sources of particular interest: a ULX, a soft X-ray source that could be a quiescent neutron star X-ray binary, and a galaxy cluster.

\section{Observations and Analysis}\label{sec:obs}

The X-ray satellite telescope \emph{XMM-Newton} observed a sample of 11 nearby late-type galaxies. These observations were taken in the period from August 2013 to January 2014 under program 72191 (PI: P.Kaaret). The observation details are given in Table~\ref{tab:observations}. All three detectors (PN, MOS1, and MOS2) that make up {\emph{part of the} European Photon Imaging Camera (EPIC) were used in Full Frame mode with the Medium filter. All data were processed using \texttt{SAS}~14.0.0\footnote{http://xmm.esac.esa.int/external/xmm\_data\_analysis/} and the event lists were created using the most recent calibration files as of 2015~July.

Several of the observations suffered from significant flaring.  In order to minimize the background, we searched for flares using the count rate in the 10$-$12~keV band and selected good time intervals (GTIs) based on the rate in each detector.  We used the standard rates of 0.4~counts~s$^{-1}$ for the PN and 0.35~counts~s$^{-1}$ for the MOS when filtering for flaring particle background in most of our observations. For a few observations, noted below, filtering at these rates left little or no exposure, so we increased the rates used for filtering in order to increase the exposure.

For each detector, we used the \texttt{evselect} task to create images in two energy bands: soft (0.2$-$2.0 keV) and hard (2.0$-$10.0 keV), with a pixel size of $4\farcs 35$ from events with FLAG = 0 and PATTERN $\leq 4$ for the PN and PATTERN $\leq 12$ for the MOS.  We then used the source detection tool \texttt{edetect\_chain} on all 6 images (2 for each detector) simultaneously.  We recorded only those sources that had a likelihood of at least 10 $(\sim 4 \sigma)$, where the likelihood is given by $-\ln p$ and $p$ is calculated as the chance probability that the detected source is a random fluctuation in the Poisson distributed background.  We compared the detected X-ray sources to an optical image of the field containing each galaxy in order to check the accuracy of the \emph{XMM-Newton} astrometry against \emph{HST} astrometry and if available \emph{Chandra} astrometry.

We only discuss those sources which fall inside or are within a $30\arcsec$ perpendicular distance of the $D_{25}$ ellipse of the galaxy. We took the $D_{25}$ ellipse dimensions from HyperLeda\footnote{http://leda.univ-lyon1.fr/}~\citep{makarov2014}.  The absorbed flux of each source was calculated from the net count rate using NASA's HEASARC WebPIMMS\footnote{https://heasarc.gsfc.nasa.gov/cgi-bin/Tools/w3pimms/w3pimms.pl} tool assuming an absorbed power law spectrum with a photon index of 2.0 and the appropriate Galactic absorption found using the HEASARC $N_{\rm H}$ column density tool\footnote{{https://heasarc.gsfc.nasa.gov/cgi-bin/Tools/w3nh/w3nh.pl}} using the \citet{Dickey1990} map.  WebPIMMS accounts for the type of camera and the filter used.  The count rate calculated by \texttt{edetect\_chain} from the source counts and exposure time was used directly.  The luminosity distance was taken to be the galaxy distance obtained from the NASA/IPAC Extragalactic Database\footnote{http://ned.ipac.caltech.edu/} (NED).  The list of detected sources is shown in Table~\ref{tab:sources}.

\begin{table*}
 \centering
 \begin{minipage}{172mm}
  \caption{X-ray sources detected with {\it XMM-Newton}.} \label{tab:sources}
  \begin{tabular}{@{}lcccccccccc@{}}

  \hline
   No.  &  Galaxy&   \multicolumn{2}{c}{Position} &&PN  &Total&&WebPIMMS & & \\
   &  & RA &  DEC  & Err & Count Rate & Count Rate &$N_{\rm H}$& Flux &Luminosity  & D$_{25}$\\

 \hline
 1. & IC 5052      & 313.070654 & $-$69.221351 & 0.12 & 0.232$\pm$0.004 & 0.394$\pm$0.006 & 4.70 & 6.59$\pm$0.11 & 287$\pm$5  &yes\\
 2. & "            & 313.002376 & $-$69.190286 & 0.48 & 0.021$\pm$0.002 & 0.036$\pm$0.002 & "    & 0.60$\pm$0.06 & 26$\pm$3    &yes\\
 3. & "            & 313.025160 & $-$69.181506 & 1.01 & 0.421$\pm$0.011 & 0.695$\pm$0.012 & "    & 12.0$\pm$0.3  & 523$\pm$13&no\\
 4. & "            & 313.043293 & $-$69.235878 & 0.96 & 0.005$\pm$0.001 & 0.009$\pm$0.001 & "    & 0.14$\pm$0.03  &6.1$\pm$1.3&no\\
 5. & "            & 313.086259 & $-$69.197722 & 9.12 & 0.045$\pm$0.008 & 0.055$\pm$0.009 & "    & 1.3$\pm$0.2     &57$\pm$9     &no\\
 6. & UGCA 442     & 355.969492 & $-$31.939676 & 1.27 & 0.005$\pm$0.003 & 0.010$\pm$0.003 & 1.14 & 0.11$\pm$0.05 &2.4$\pm$0.1&yes\\
 7. & NGC 4605     & 190.054019 & +61.597501 	& 2.62 & 0.047$\pm$0.014 & 0.050$\pm$0.014 & 1.43 & 1.0$\pm$0.3     &36$\pm$11   &yes\\
 8. & "            & 189.974712 & +61.640053 	& 1.68 & 0.017$\pm$0.013 & 0.022$\pm$0.013 & "    & 0.4$\pm$0.3     &14$\pm$11   &no\\
 9. & ESO 154-G023 &  44.267194 & $-$54.546250 & 0.47 & 0.053$\pm$0.005 & 0.086$\pm$0.006 & 1.87 & 1.21$\pm$0.12 &48$\pm$5     &yes\\
10. & "            &  44.250823 & $-$54.540720 & 2.61 & 0.005$\pm$0.002 & 0.008$\pm$0.002 & "    & 0.12$\pm$0.05 &4.8$\pm$1.9 &yes\\
11. & "            &  44.217217 & $-$54.562757 & 1.21 & 0.008$\pm$0.003 & 0.015$\pm$0.003 & "    & 0.19$\pm$0.06 &8$\pm$2       &yes\\
12. & "            &  44.172706 & $-$54.599862 & 1.93 & 0.012$\pm$0.003 & 0.012$\pm$0.003 & "    & 0.28$\pm$0.07 &11$\pm$3      &yes\\
13. & IC 4662      & 266.788302 & $-$64.636824 & 0.87 & 0.022$\pm$0.002 & 0.033$\pm$0.003 & 6.22 & 0.69$\pm$0.07 &4.9$\pm$0.5 &yes\\
14. & ESO 383-G087 & 207.331072 & $-$36.073609 & 1.18 & 0.008$\pm$0.002 & 0.011$\pm$0.002 & 4.97 & 0.22$\pm$0.05 &3.1$\pm$0.7  &yes\\
15. & "            & 207.338584 & $-$36.080529 & 0.99 & 0.012$\pm$0.002 & 0.017$\pm$0.002 & "    & 0.34$\pm$0.06 &4.8$\pm$0.9  &no\\
16. & NGC 5264     & 205.407050 & $-$29.916541 & 1.11 & 0.013$\pm$0.002 & 0.021$\pm$0.003 & 3.79 & 0.35$\pm$0.06 &8.5$\pm$1.5  &yes\\
17. & NGC 1311     &  50.014727 & $-$52.200464 & 1.53 & 0.042$\pm$0.016 & 0.046$\pm$0.016 & 2.50 & 1.0$\pm$0.4     &36$\pm$14    &yes\\
18. & "            &  50.007126 & $-$52.179953 & 1.44 & 0.008$\pm$0.002 & 0.011$\pm$0.003 & "    & 0.21$\pm$0.06 &8$\pm$2        &no\\
\hline
\end{tabular}

\textbf{Notes.} Columns 1-5: source number, galaxy name, source coordinates in degrees (J2000), and positional statistical errors in arcseconds. Columns 6-7: count rates in the PN and the combined rates for all three detectors in counts s$^{-1}$. Column 8: Galactic absorption column density in $10^{20}$ cm$^{-2}$. Column 9: unabsorbed PN fluxes (assuming a power law with $\Gamma=2.0$) in $10^{-13}$ erg cm$^{-2}$ s$^{-1}$ for the energy band 0.2$-$10.0~keV. Column 10: unabsorbed Luminosity in the 0.2$-$10.0~keV band in units of $10^{37}$~erg~s$^{-1}$, assuming the distance to the galaxy as luminosity distance. Column 11 denotes whether the source is inside the $D_{25}$ ellipse boundary of the galaxy (`yes') or within $30\arcsec$ of the $D_{25}$ ellipse (`no'). \\
\end{minipage}
\end{table*}

Using the \texttt{especget} tool, spectra and responses were obtained for sources in Table~\ref{tab:sources} with more than 200 total EPIC counts. The resulting spectra were binned with a minimum of 16 counts in each bin. The source regions were selected as circles with a radius of $36\arcsec$, except where the radius was decreased to avoid overlapping a nearby source as noted below. The background regions were chosen to be circular with no detected sources, with a radius of at least $100\arcsec$, and were located on the same chip as the detected source. The spectral fitting program XSPEC~v12.8.1~\citep{arnaud1996}  was used to fit models using $\chi^2$-statistics. The extracted spectra for the PN and both MOS detectors were simultaneously fitted with a power law model subject to photo-electric absorption (with Wisconsin cross-sections; \citealt{b20}). We chose a power law as our initial model as we expect these objects to be X-ray binaries, which typically exhibit power law spectra. The results of power law fitting are given in Table~\ref{tab:c} and shown in Fig~\ref{fig:3spectra}.  We find that this model adequately describe these spectra. The fluxes from XSPEC are larger than those from WebPIMMS since we used only the Galactic absorption coefficient for the WebPIMMS calculation without accounting for intrinsic absorption.

In the following, we provide notes on the analysis and results for each individual galaxy. We make note of any known X-ray or optical counterparts with the goal to cross check the astrometry across instruments.

\begin{figure*}
\includegraphics[width=0.5\textwidth]{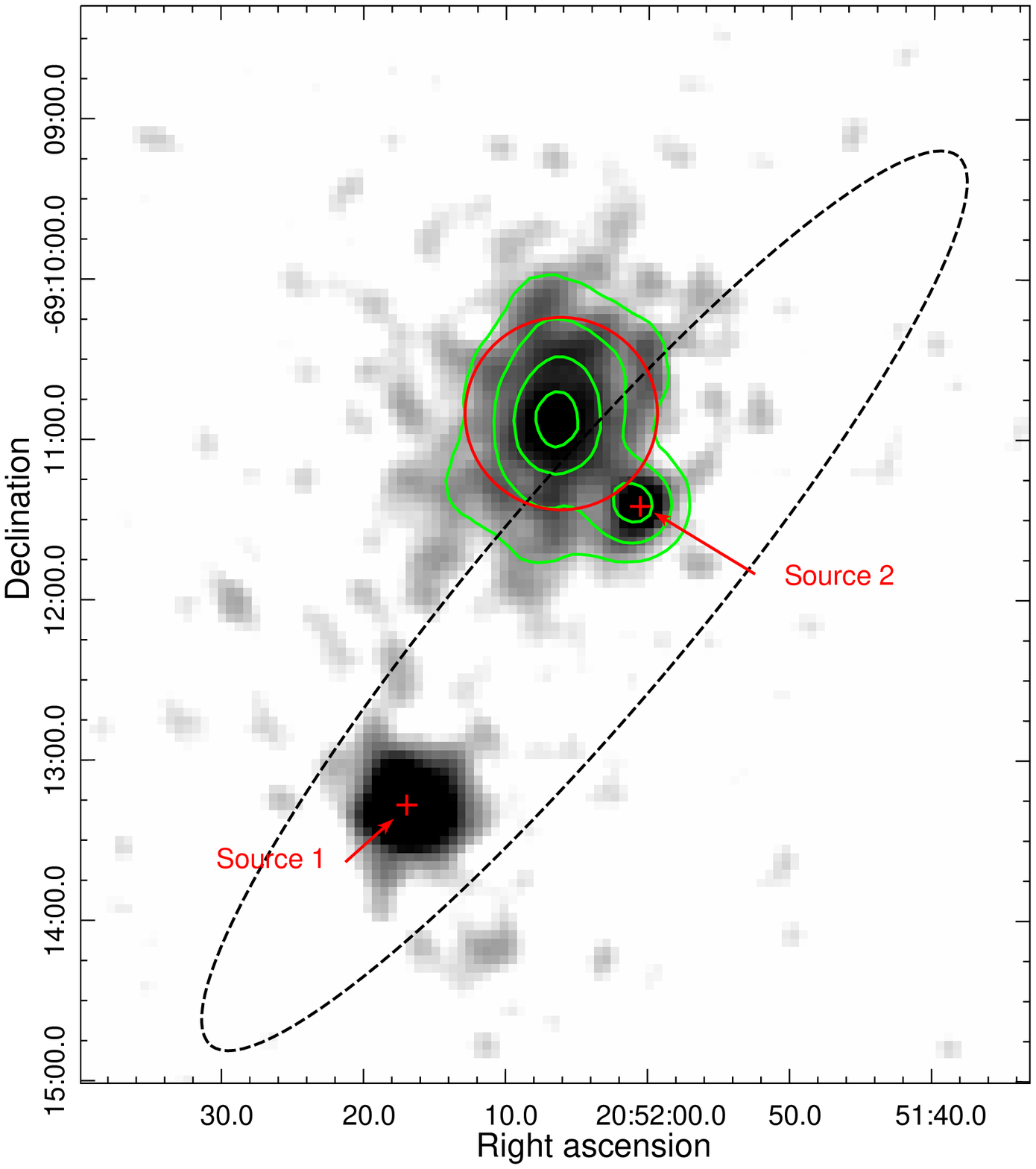}\hfil
\includegraphics[width=0.49\textwidth]{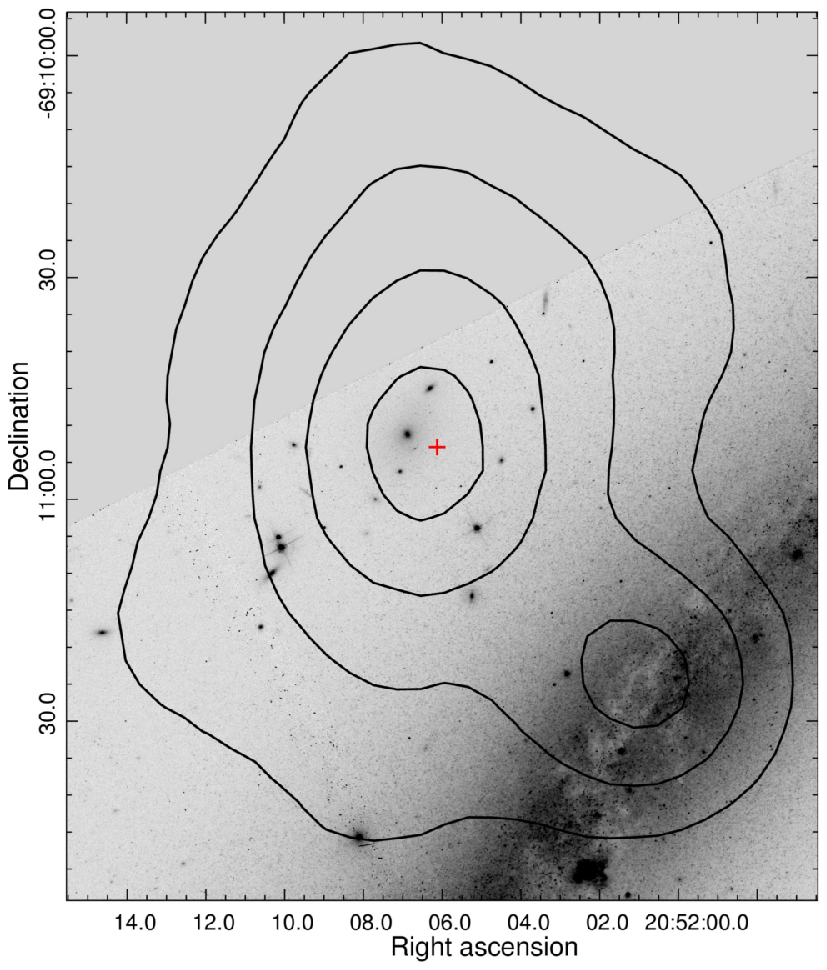}
\caption{\emph{Left}: X-ray image of IC~5052.  Sources 1 and 2 are marked as red crosses, while the extraction region used for the extended source is marked as a red circle.  X-ray contours are shown in green for the extended source and the D$_{25}$ ellipse is shown as a black dashed line. \emph{Right}: \emph{HST} image of the field containing the extended source XMMU J205206.0$-$691316 obtained with the ACS/WFC using the F606W filter. The red cross shows the centroid of the X-ray source and the black contours (identical to the green contours in figure to the left) represent the extent of the X-ray emission.}
\label{fig:xray_pgc65603}
\end{figure*}

\begin{figure*}
\includegraphics[width=\textwidth]{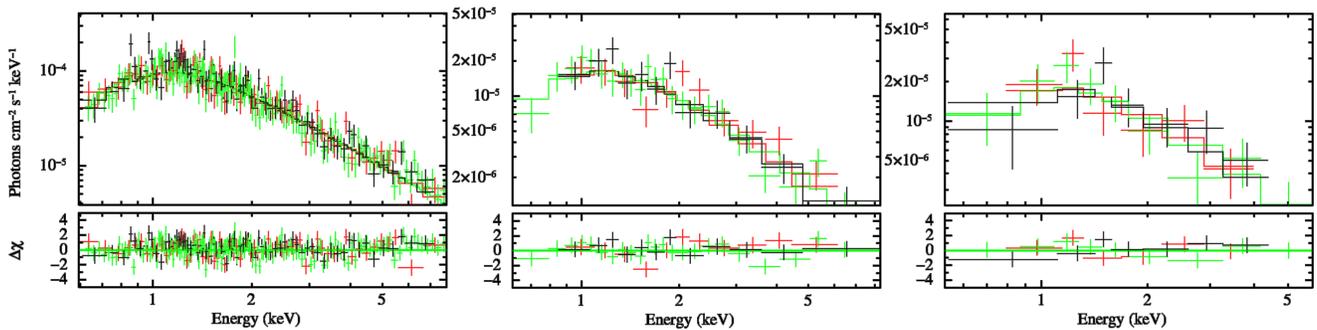}
\caption{Unfolded spectra from sources in Table~\ref{tab:c}. Green, black and red data points are PN, MOS1, and MOS2 data, respectively. \emph{Left}: Power law spectrum of Source~1 from IC~5052. \emph{Middle}: Power law spectrum of Source~2 from IC~5052. \emph{Right}: Power law spectrum of Source~9 from ESO~$154-$G$023$.}
\label{fig:3spectra}
\end{figure*}

\begin{table*}
 \centering
 \begin{minipage}{130mm}
  \caption{Spectral parameters of bright bources with power law fitting} \label{tab:c}

  \begin{tabular}{@{}lccccccc@{}}
  \hline
 No.&Galaxy& $kT_{\rm in}$  & $\Gamma$	&  $N_{\rm H}$	&  $f_{\rm X}$ &  $L_{\rm X}$ & $\chi^2/$d.o.f. \\
 \hline
 1. & IC 5052		& 	$-$						& $2.06\pm0.08$ 	&$0.45\pm0.04$	&$16.6^{+1.2}_{-1.0}$	&$7.2\pm0.5$	&287.03/303\\
 	& "				& $0.45^{+0.15}_{-0.20}$ 	& $1.7^{+0.3}_{-0.5}$&	$0.42^{+0.12}_{-0.08}$& $14.3^{+5.0}_{-1.8}$	& $6.2^{+2.1}_{-0.8}$	& 279.13/301\\
 2. & " 			&   $-$						& $2.0\pm0.2$ 	&$0.44\pm0.12$	& $2.8^{+0.6}_{-0.4}$	&$1.2^{+0.3}_{-0.2}$	&38.78/43\\
 9. & ESO 154$-$G023	& $-$					& $1.8_{-0.5}^{+0.7}$ & 0.4$\pm$0.2	& $3.0^{+2.0}_{-0.6}$ & $1.2^{+0.8}_{-0.3}$& 15.53/19\\
 \hline
\end{tabular}
\textbf{Notes.} $kT_{\rm in}$ is the temperature at the inner disk radius in keV. $\Gamma$ is the power-law photon index. $N_{\rm H}$ is the absorption column density in $10^{22}$ cm$^{-2}$. The energy range over which the flux and luminosities are calculated is 0.2$-$10.0~keV. $f_{\rm X}$ is the unabsorbed flux in units of  $10^{-13}$ erg cm$^{-2}$ s$^{-1}$. $L_{\rm X}$ is the unabsorbed luminosity in units of $10^{39}$ erg  s$^{-1}$, taking the  distance to the galaxy as luminosity distance. Errors are quoted at 90\% confidence level.
\end{minipage}
\end{table*}

\subsection{\textit{IC 5052}}

The \emph{XMM-Newton} observation of IC~5052 showed no significant background flaring, and so no flare filtering was applied to the EPIC data. We compared the 100 detected X-ray sources in the \emph{XMM-Newton} field of view to an optical image and found an X-ray source to be within $0\farcs 6$ of the star CD-69~1954 (source: SIMBAD\footnote{http://simbad.u-strasbg.fr/simbad/}). The field of a \emph{Chandra} observation partially overlaps the \emph{XMM-Newton} field, but does not include the galaxy.  Four \emph{Chandra} sources fall within the \emph{XMM-Newton} field. One \emph{Chandra} source with good location accuracy matched within $1.9\arcsec$ of an \emph{XMM-Newton} source. The others have large positional errors or were not detected with \emph{XMM-Newton}. Based on the positional errors of the coincident \emph{Chandra} source and the catalogued star CD-69~1954 with respective \emph{XMM-Newton} sources, we conclude that the \emph{XMM-Newton} astrometry is accurate within $2\arcsec$.  An \emph{HST} image of IC~5052, taken with the ACS/WFC using the F606W filter on 2003~December~14 (PI: de Jong), was compared with the \emph{XMM-Newton} image, matching astrometry with \emph{XMM-Newton} sources within centroid positional errors, $<1.5\arcsec$.

An X-ray image of the galaxy is shown in Figure~\ref{fig:xray_pgc65603}. Two sources are inside the D$_{25}$ ellipse and three are overlapping, with their centroids outside the D$_{25}$ region. One of the sources outside the galaxy is extended with a measured extent of $31.8\arcsec$. The source extent is defined as the Gaussian sigma or beta model core radius. This source is discussed in section~\ref{sec:cluster}.

The two sources within the galaxy appear to be point like, each with an extent of less than $6\arcsec$. The flux of source 1 calculated using WebPIMMS corresponds to an unabsorbed luminosity of (2.87$\pm$0.05) $\times$ 10$^{39}$ erg s$^{-1}$. Thus, we classify it as a ULX. This source is discussed further in section~\ref{sec:ulx}.

The extraction region for source~2 was chosen to have a radius of $19.7\arcsec$ to avoid overlapping the extended source. Spectral fitting in the 0.2$-$10~keV band provided a power law photon index of $2.0\pm0.2$ and an unabsorbed luminosity of $(1.09 \pm 0.17) \times 10^{39}$~erg~s$^{-1}$ (see Table~\ref{tab:c} for more details).

\subsection{\textit{UGCA 442}}

There was significant flaring in the PN, reducing its usable live time from 19.9~ks to 7.8~ks. The MOS suffered much less flaring, resulting in 18.4~ks of good observation time. An X-ray source was found to be within $2\farcs 4$ of the star CD-32~17640, which is within the positional error of \emph{XMM-Newton}. There was only one source detected inside the D$_{25}$ ellipse. It appears to be a point source with fewer than 100 counts. Assuming an absorbed power law model with $\Gamma=2.0$, we calculated an unabsorbed luminosity of $(2.4 \pm 0.1) \times 10^{37}$~erg~s$^{-1}$.

\subsection{\textit{NGC 4605}}

There was strong flaring in the PN which required us to increase the PN rate filter to 1.0~counts~s$^{-1}$ and thereby reducing its usable observation time to only 500 sec.  An X-ray source was found to be within $1\farcs 1$ of QSO B1236.6+6200. There was also a source detected inside the D$_{25}$ ellipse, with another source extending partially into the region. Both appear to be point sources with fewer than 50 counts and likelihoods above 13. We calculated, for the source inside the D$_{25}$ ellipse, an unabsorbed luminosity of $(3.6 \pm 1.1) \times 10^{38}$~erg~s$^{-1}$.

\subsection{\textit{ESO 154-G023}}

There was large flaring in all three detectors, thus we increased the filtering rates to 1.0~counts~s$^{-1}$ for the PN and 0.5~counts~s$^{-1}$ for the MOS.  This resulted in usable observation times of only 2.8, 6.7, and 6.8~ks for the PN, MOS1, and MOS2, respectively.  An X-ray source was found to be within $0\farcs 5$ of ultraviolet source GALEXASC J025627.19$-$543358.3. There were four sources detected inside the D$_{25}$ ellipse. All are point sources. Source~9 is the brightest with 354 net counts and has an unabsorbed luminosity of $(4.8 \pm 0.5) \times 10^{38}$~erg~s$^{-1}$ in the 0.2$-$10.0~keV energy band calculated the WebPIMMS tool.  The fit of a power-law model to the spectrum of Source~9 in the 0.6$-$5.0~keV energy band gave a photon index of $1.8_{-0.5}^{+0.7}$ and an unabsorbed luminosity of $1.0^{+0.9}_{-0.3} \times 10^{39}$~erg~s$^{-1}$ (full details in Table~\ref{tab:c}).

\subsection{\textit{IC 4662}}

There was no significant flaring in any of the CCDs.  An X-ray source was found to be within $2\farcs 0$ of the star CD-64 1144.  There was only one source (Source 13) detected inside the D$_{25}$ ellipse.  It appears to be a point source with more than 250 counts, all detected below 2.0~keV. This source is discussed in section~\ref{sec:softsource}.

\subsection{\textit{ESO 383-G087}}

There was strong flaring in the PN, reducing its usable observation time to 7.1~ks after increasing the rate filtering threshold to 0.75~counts~s$^{-1}$.  An X-ray source was found to be within $0\farcs 9$ of the star CD-35 9030. There was one source detected inside the D$_{25}$ ellipse with another source extending partially into the region. Both are point sources with close to 100 and 150 counts respectively. We calculated, for the source inside the D$_{25}$ ellipse, an $L_{\rm X}$ value of $(3.1 \pm 0.7) \times 10^{37}$~erg~s$^{-1}$.  The other source had an unabsorbed luminosity of $(4.8 \pm 0.9) \times 10^{37}$~erg~s$^{-1}$.

\subsection{\textit{NGC 5264}}

There was one small window of significant flaring in each CCD. An X-ray source was found to be within $0\farcs 2$ of the star HD~119136 (source: SIMBAD). There was only one source detected inside the D$_{25}$ ellipse. It appears to be a point source with almost 100 counts and a likelihood of 21.8. We calculated an $L_{\rm X}$ value of  $(8.5 \pm 1.5) \times 10^{37}$~erg~s$^{-1}$.

\subsection{\textit{NGC 1311}}

There was strong flaring in all three detectors and we increased the PN rate filtering threshold to 1.0~counts~s$^{-1}$.  An X-ray source was found to be within $1\farcs 9$ of the ultraviolet source GALEXASC J032052.52$-$520803.7. There was one source detected inside the D$_{25}$ ellipse, with another source extending partially into the region. Both are point sources with fewer than 100 counts.  For the source inside the D$_{25}$ ellipse (Source~17), we calculated an unabsorbed X-ray luminosity of $(3.6 \pm 1.4) \times 10^{38}$~erg~s$^{-1}$. The other source had an unabsorbed luminosity of $(8 \pm 2) \times 10^{37}$~erg~s$^{-1}$.

\section{Results and Discussion}
Using \emph{XMM-Newton} data, we found 12 X-ray sources within 8 galaxies in a sample of 11 galaxies observed for this study. The photometric X-ray luminosities of the sources vary from $0.024-5 \times 10^{39}~{\rm erg s}^{-1}$ in the $0.2-10$~keV energy range. Table~\ref{tab:sources} further summarizes the results of Section~\ref{sec:obs}. In the following sections, we present more detailed results for three objects found within our observations; a ULX in IC~5052, a soft X-ray source in IC~4662, and a galaxy cluster near IC~5052.

\subsection{Survey results and a ULX in \textit{IC 5052}}
\label{sec:ulx}

\begin{figure}
\includegraphics[width=0.49\textwidth]{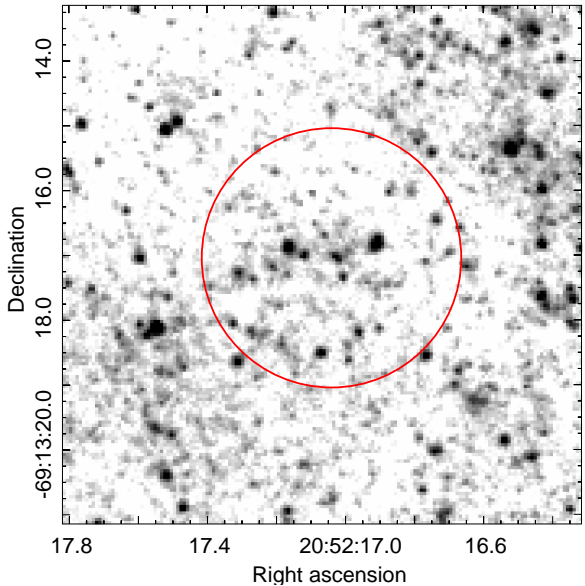}
\caption{\emph{HST} image of IC~5052~X-1 obtained with the ACS/WFC using the F606W filter. The ULX is contained within the $2\arcsec$ radius of the positional error circle (red). We are unable to determine a unique optical counterpart to the X-ray source.}
\label{fig:hst_ulx}
\end{figure}

The number of high-mass X-ray binaries found in a galaxy is proportional to the host's star formation rate (SFR; \citealt{grimm2003}).  Using the X-ray luminosity function (XLF) measured for normal metallicity galaxies \citep{b5} and taking the sum total of the SFRs for all of the galaxies in our sample (see Table~\ref{tab:observations}), we can detemine the expected number of sources, $N$, above a minimum luminosity $L_{\rm min}$. Following \citet{b6} and taking $L_{\rm min} = 2\times 10^{39}$~erg~s$^{-1}$, we find $N = 0.11$. We found one ULX in our sample, which is consistent.

The brightest source in IC~5052 is the only ULX identified in this survey. The ULX has a total of 5400 net counts from all three detectors. We name the source XMMU J205216.9$-$691316 = IC 5052 X-1. We searched for an optical counterpart within the ACS/WFC HST image (see Figure~\ref{fig:hst_ulx}). No unique optical counterpart can be identified. This may be due to absorption effects given that we see the galaxy edge-on. 

We extracted spectra for IC~5052~X-1 as described in Section~\ref{sec:obs} and fitted an absorbed power law (see Table~\ref{tab:c}). Our best fitting parameters are a photon index of $\Gamma = 2.06\pm 0.08$ and a column density of $N_{\rm H} = 0.45\pm 0.04 \times 10^{22} {\rm cm}^{-2}$, giving a $\chi^2/{\rm d.o.f.} = 287.03/303$. If we introduce a disk blackbody component to the model we find $\Gamma = 1.7^{+0.3}_{-0.5}$, $kT_{\rm in} = 0.45^{+0.15}_{-0.20}~{\rm keV}$, $N_{\rm H} = 0.42^{+0.12}_{-0.08} \times 10^{22} {\rm cm}^{-2}$, and $\chi^2/{\rm d.o.f.} = 279.13/301$. The F-test gives a probability of 0.013 which indicates that it is reasonable to add this second component to our model, though it is not a strong improvement to our fitting. A single component model of a disk blackbody on its own without a power law resulted in a poor fit with $\chi^2/{\rm d.o.f.} = 406.10/303$.

\cite{winter2006} discovered a bimodality in ULX temperatures where the population of ULXs tended to cluster around disk temperatures of $kT_{\rm in} = 0.1$ and $1.0$~keV, leaving few with disc temperatures of about $0.5$~keV. They suggested that this bimodality arises from a difference in the nature of the compact object. For lower disc temperatures, the object is a candidate IMBH. For higher temperatures, a candidate stellar mass BH is proposed for the object. IC~5052~X-1 lies at the high temperature end of the IMBH candidate range.

\begin{figure}
\includegraphics[width=2.25in,angle=270]{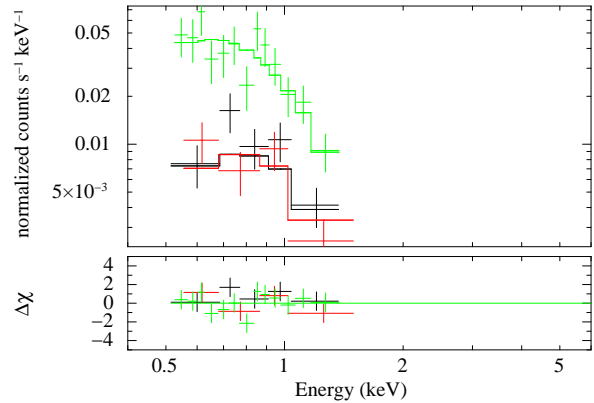}
\caption{Spectrum of the soft X-ray source IC 4662 X-1 fitted with a blackbody model. The colors are the same as Figure~\ref{fig:3spectra}.}
\label{fig:soft}
\end{figure}

\begin{table}
 \begin{center}
  \caption{Best fitting parameters for the spectrum of the soft X-ray source IC 4662 X-1 and the ULX IC~5052~X-1.} \label{tab:soft}
  \begin{tabular}{@{}lcc@{}}
  \hline
  			  &	\multicolumn{2}{c}{IC~4662 X-1} 	\\
  			  & \multicolumn{2}{c}{(WABS)}			\\
              & Power Law       & Blackbody       	\\
$N_{\rm H}$   & 6.22            & 6.22            	\\
$\Gamma$      & 3.4$\pm$0.4     & ---             	\\
$kT$          &   ---           & 0.166$\pm$0.015 	\\
$f_{\rm X}$   & 5.8$\pm$0.6     & 5.2$\pm$0.5     	\\
$L_{\rm X}$   & 4.1$\pm$0.4     & 3.7$\pm$0.4     	\\
$\chi^2$/d.o.f.  & 24.51/20     & 19.67/20     	\\
\hline
\end{tabular}
\end{center}
\textbf{Notes.} Row 1-2: XSPEC absorption/continuum models; Rows 3-8 : Model parameters: $N_{\rm H}$ is the absorption column density in $10^{20}$ cm$^{-2}$, $\Gamma$ is the power-law photon index, $kT$ is the temperature in keV, $f_{\rm X}$ is the unabsorbed flux in units of $10^{-14}$ erg cm$^{-2}$ s$^{-1}$ for the range 0.5$-$2.0~keV for IC~4662~X-1 and in units of $10^{-12}$ erg cm$^{-2}$ s$^{-1}$ for the range 0.2$-$10.0~keV for IC~5052~X-1. $L_{\rm X}$ is the unabsorbed luminosity in the 0.5$-$2.0~keV band in units of $10^{37}$~erg~s$^{-1}$ for IC~4662~X-1 and in the 0.2$-$10.0~keV band in units of $10^{39}$~erg~s$^{-1}$ for IC~5052~X-1.  Errors are quoted at the 90\% confidence level. For the powerlaw and blackbody models of IC~4662~X-1 and one of the TBABS components of IC~5052~X-1, $N_{\rm H}$ was set to the column density along the line of sight in the Milky Way.\\
\end{table}

\subsection{A soft source in \textit{IC 4662}}
\label{sec:softsource}

The spectrum of the X-ray source found in IC~4662 contained more than 250 counts, but extended only up to 2~keV, indicating a soft source. We name the source XMMU J174709.9$-$643812 = IC 4662 X-1.  Spectral fitting was done in the 0.5$-$2.0~keV band with power law and blackbody models, each subjected to the WABS interstellar absorption model.  The spectrum is shown in Figure~\ref{fig:soft} and the fit results are presented in Table~\ref{tab:soft}.  The powerlaw and blackbody models provided adequate fits with no absorption beyond that along the line of sight in the Milky Way.  

We examined an \emph{HST} image taken on 2005~October~15 with the ACS/HRC (PI: Vacca) that covered the region, but there is no obvious single counterpart such as a bright, foreground star. The low temperature of the blackbody fit, $kT \sim 0.2$~keV, could suggest a neutron star in quiescence.  The Galactic latitude is $b$ = $-$17.8. A distance of 10~kpc would give a luminosity of $6 \times 10^{32}$~erg~s$^{-1}$, consistent with known quiescent X-ray binaries containing a neutron star~\citep{gendre2003}. Using the \texttt{bbodyrad} model in XSPEC to fit the spectrum, we found a normalization of $R_{\rm km}^2 / D_{\rm 10kpc}^2 = 10.5^{+6.3}_{-4.0}$ which corresponds to a neutron star of radius 3.2~km for a distance of 10~kpc.

Super-soft sources produce most of their emission below 1~keV and have typical luminosities of $10^{36-38}$~erg~s$^{-1}$. Their spectra are usually fitted with a blackbody model with a temperature $<100$~eV \citep{b2}. Quasi-soft sources (QSSs) have similar properties, but their spectra show higher temperatures, in the range 100$-$350~eV, when fitted as a blackbody. When fitted as a powerlaw, they have steep photon indices around 3 or higher.  Our measured values are consistent with these ranges. Also, assuming the source lies within IC~4662, both models give $L_{\rm X} \sim 4 \times 10^{37}$~erg~s$^{-1}$. Thus, our source could be classified as a QSS. The physical nature of QSSs is still unknown, and \cite{b2} suggests three possibilities: nuclear burning white dwarfs, intermediate mass black holes, and supernova remnants.

\begin{figure}
\includegraphics[width=2.25in,angle=270]{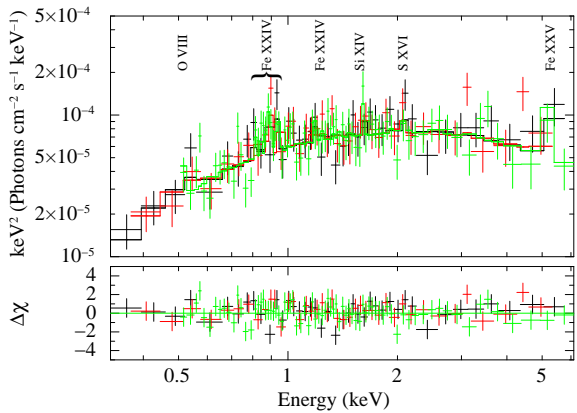}
\caption{Unfolded X-ray spectrum of the galaxy cluster XMMU J205206.0$-$691316 fitted with an absorbed APEC model (see parameter values in Table~\ref{tab:clusterxspec}; colors are the same as Figure~\ref{fig:3spectra}). We have added labels for emission lines that are expected to be most prominent given the temperature and redshift from the best fit APEC model parameters.}
\label{fig:clusterxspec}
\end{figure}

\begin{table}
\begin{center}
\caption{Best fitting parameters for the X-ray spectrum of the galaxy cluster XMMU J205206.0$-$691316} \label{tab:clusterxspec}
\begin{tabular}{@{}lccc@{}}
\hline
             & APEC          & MEKAL         \\
$N_{\rm H}$  & 5.80$\pm$0.02 & 5.79$\pm$0.02 \\
$kT$         & 3.7$\pm$0.4   & 3.6$\pm$0.05  \\
Abundance    & 0.35$\pm$0.18 & 0.5$\pm$0.3   \\
Redshift     & 0.25$\pm$0.02 & 0.24$\pm$0.02 \\
Norm         & 3.6$\pm$0.4   & 3.5$\pm$0.4   \\
$f_{\rm X}$  & 2.94$\pm$0.10 & 2.95$\pm$0.10 \\
$L_{\rm X}$  & 6.1$\pm$0.2   & 5.61$\pm$0.19 \\
$\chi^2$/d.o.f. & 138.49/154 & 138.94/154    \\
\hline
\end{tabular}
\end{center}
\textbf{Notes.} Row 1 : XSPEC model. Rows 2-9 : Model parameters: $N_{\rm H}$ is the absorption column density in units of $10^{20}$ cm$^{-2}$, $kT$ is the temperature in keV, Norm is the normalization parameter in units of $10^{-4}$ cm$^{-5}$, $f_{\rm X}$ is the unabsorbed flux in units of $10^{-13}$ erg cm$^{-2}$ s$^{-1}$ for the energy range 0.5$-$7.0~keV, $L_{\rm X}$ is the unabsorbed luminosity in the 0.5$-$7.0~keV band in units of $10^{43}$~erg~s$^{-1}$.  Errors are quoted at the 90\% confidence level.  Luminosity figures are calculated using \citet{b21}. \\
\end{table}

\subsection{A Galaxy Cluster near \textit{IC 5052}}
\label{sec:cluster}

We found an extended X-ray source near the galaxy IC~5052. Hereafter, we refer to the source as XMMU J205206.0$-$691316 or J2052. The flux from J2052 did not show any time variation over the observation. We overlayed the X-ray contours of this source on an \emph{HST} image taken using the ACS/WFC with the F606W filter (2003-12-14; PI: de~Jong), see Figure~\ref{fig:xray_pgc65603}, and found that it coincides with a group of galaxies. This proposed galaxy cluster has not been previously catalogued.  

Identification of the source as a galaxy cluster suggests that the X-ray emission is due to a hot, thermal plasma. We fitted the spectrum using the thermal APEC model \citep{apec}, describing emission from collisionally-ionised diffuse gas calculated using the atomic database (ATOMDB) code\footnote{http://atomdb.org/}, subject to absorption modeled using WABS. We allowed the cluster redshift and abundance to vary. The best fitting spectrum is shown in Figure~\ref{fig:clusterxspec} and the parameters are given in Table~\ref{tab:clusterxspec}.  The fitted redshift is determined from emission lines present in the spectrum, the strongest of which are a cluster of iron lines near 1~keV.  We also fitted the spectrum using the MEKAL model \citep{mekal} and found very similar results.

Based on the fitted redshift values from the APEC ($z$ = 0.25$\pm$0.02) and MEKAL ($z$ = 0.24$\pm$0.02) models, we determined the comoving radial distance ($d_{\rm r}$), luminosity distance ($d_L$) and angular size distance ($d_\theta$) to the cluster using Ned Wright's Javascript Cosmology Calculator \citep{b21} using values of Hubble's constant $H_0$ = 67$\pm$1.2 km s$^{-1}$ Mpc$^{-1}$ \citep{b22}, matter density $\Omega_{\rm M}$ = 0.286 and vacuum density $\Omega_{\rm vac}$ = 0.714.

For the best fitted redshift, $z$ = 0.25, from the APEC model, $d_{\rm r}$ = 1056.0 Mpc, $d_L$ = 1320.0 Mpc and $d_\theta$ = 4.096 kpc/$\arcsec$. For this distance, the cluster's radius of 31.8$\arcsec$ gives a linear radius $r$ = 130 kpc. From the luminosity distance, we find an X-ray luminosity of $(6.1 \pm 0.2) \times 10^{43}$~erg~s$^{-1}$ for the energy range 0.5$-$7.0~keV. In the $0.2-10$~keV range we calculate a luminosity of $(7.3 \pm 0.5) \times 10^{43}$~erg~s$^{-1}$. The APEC normalization parameter is defined as:

\begin{equation}
{\rm Normalization} = \frac{10^{-14}}{4\pi d_{\rm r}^2}\int n_{\rm e} n_{\rm H}dV,
\end{equation}

\noindent where $n_{\rm e}$ and $n_{\rm H}$ are, respectively, the electron and hydrogen number densities in cm$^{-3}$ and $V=\int dV$ is the X-ray emitting gas volume. Assuming it to be spherical, $V=(4/3)\pi r^3 = 2.7 \times 10^{71}$~cm$^3$. From the normalization of the APEC fit, assuming $n_{\rm e} = n_{\rm H}$, we find $n_{\rm H}$ = 0.0042 cm$^{-3}$ for the X-ray emitting plasma. Additionally, by using $M = n_{\rm H} m_{\rm H} V$, where $m_{\rm H}$ is the hydrogen mass value, we can estimate the mass of the X-ray emitting gas to be $M  = 9.6 \times 10^{11} {\rm M}_{\sun}$. If we take into account that the cluster's extent radius of 31.8$\arcsec$ is a characteristic profile radius that encloses 71~per~cent of the mass, then the X-ray emitting gas may have a mass as large as $M  = 1.4 \times 10^{12} {\rm M}_{\sun}$. Since the $R_{500}$ of the cluster is likely larger than the radius of the extraction region used, the X-ray luminosity and mass calculations likely underestimate the true values.

\section{Conclusions}\label{sec:con}

Surveys have been an essential part of astronomy since its beginning.  They are necessary to identify new classes of objects and to accumulate samples of known classes.  We conducted a relatively modest X-ray survey of 11 nearby late-type galaxies with the primary goal being the identification of new ULXs.  We found one ULX, which is located in IC 5052.  This new ULX lies at the high temperature end of the sub-class of ULXs with spectra fit with cool, $kT \sim 0.1$~keV thermal emission.  Further studies, particularly on its spectral variability, could help elucidate the physical nature of this source and might help to shed light on the nature of that sub-class of ULX.

Beyond our search for ULXs, we identified a new soft X-ray source coincident with IC 4662.  This source may be a quasi-soft source in that galaxy or a quiescent neutron-star in the Milky Way.  An improved localization enabled by a future \emph{Chandra} observation, followed up by the identification of its optical counterpart could allow us to distinguish between these two possibilities.  We also discovered a new cluster of galaxies located near IC 5052.  Fitting the X-ray spectrum suggests a redshift of $z = 0.25 \pm 0.02$.

\section*{acknowledgements}

K.~C. acknowledges support for an S.~N.~Bose Scholarship.

\label{lastpage}
\end{document}